\documentclass{pkas}

\def\year{2023} 
\def\month{Dec} 
\def\volume{38} 
\def\issue{3} 
\def\beginpage{1} 
\def\endpage{13} 
\def\received{July 07, 2023} 
\def\accepted{October 29, 2023} 
\date{Received \received ; accepted \accepted}

\usepackage{amsmath,amssymb,latexsym}
\usepackage{natbib}
\usepackage[subrefformat=parens]{subcaption}
\usepackage{enumitem}
\usepackage{aas_macros}
\setlist[itemize]{align=parleft} 
\setlist[description]{labelindent=\parindent, itemsep=0pt, parsep=0pt, topsep=0pt, partopsep=0pt}

\newcommand{\vrot}{v_\mathrm{rot}}
\def\arcdeg{^{\circ}}

\def\farcs{\hbox{$.\!\!^{\prime\prime}$}}

\def\mJB{{\rm mJy\,beam^{-1}}}

\def\kms{{\rm km\,s^{-1}}}
\def\Msun{M_{\odot}}

\def\II{{\rm I\hspace{-0.1em}I}}

\def\aap{A\&A}

\title{
Spectral Line Analysis/Modeling (SLAM) I: pvanalysis
}

\author[1]{Yusuke Aso}
\author[2]{Jinshi Sai (Insa Choi)}

\affil[1]{Korea Astronomy and Space Science Institute, 776 Daedeok-daero, Yuseong-gu, Daejeon 34055, Republic of Korea}
\affil[2]{Academia Sinica Institute of Astronomy and Astrophysics, 11F of Astronomy-Mathematics Building, AS/NTU, No.1, Sec. 4, Roosevelt Rd, Taipei 10617, Taiwan}

\def\corrauthor{
Y. Aso
}

\def\runningauthor{
Y. Aso
}

\def\runningtitle{
Pvanalysis in SLAM
}

\def\keywords{
Circumstellar disks (235) --- Circumstellar envelopes (237)
}

\def\abstracttext{
Line observations of young stellar objects (YSOs) at (sub)millimeter wavelengths provide essential information of gas kinematics in star and planet forming environments. For Class 0 and I YSOs, identification of Keplerian rotation is of particular interest, because it reveals presence of rotationally-supported disks that are still being embedded in infalling envelopes and enables us to dynamically measure the protostellar mass. We have developed a python library \texttt{SLAM} (Spectral Line Analysis/Modeling) with a primary focus on analyses of emission line data at (sub)millimeter wavelengths. Here, we present an overview of the \texttt{pvanalysis} tool from \texttt{SLAM}, which is designed to identify Keplerian rotation of a disk and measure the dynamical mass of a central object using a position-velocity (PV) diagram of emission line data. The advantage of this tool is that it analyzes observational features of given data and thus requires few computational time and parameter assumptions, in contrast to detailed radiative transfer modelings. In this article, we introduce the basic concept and usage of this tool, present an application to observational data, and discuss remaining caveats.
}

\begin{document}
\pkashead 

\section{Introduction} \label{sec:intro}
The circumstellar disk is a key structure in low mass star formation because planet formation is expected to occur in the disks where material is supported by rotation against the gravity of the central star in a young stellar object (YSO). The rotation balancing with the gravity is called Keplerian rotation and is expressed by the rotational velocity of 
\begin{align}
    v_\mathrm{rot} (r) = \sqrt{\frac{G M_\ast}{r}},
    \label{eq:vkep}
\end{align}
where $r$ is the radius, $G$ is the gravitational constant, and $M_\ast$ is the central stellar mass. Disks around YSOs are expected to exhibit Keplerian rotation when the gravity of the central star is a dominant force determining gas motion. Keplerian disks were identified first around YSOs in the Class $\II$ phase, called protoplanetary disks \citep{Koerner:1993aa, Dutrey:1998aa, Guilloteau:1998aa}. In this evolutionary phase, a YSO simply consists of a central star and a disk, and thus Keplerian rotation of the disks can be identified relatively easily through molecular line observations at millimeter and submillimeter wavelengths.

Disks have been also identified around YSOs in earlier phases, Class 0/I or protostellar phase \citep[e.g.,][]{Lee:2010aa, Tobin:2012aa,ohas14}. In this evolutionary phase, a YSO, i.e., a protostellar system consists of a central protostar, a disk, an envelope surrounding the disk, and an outflow driven by the rotation in the disk. Because envelopes and outflows have different kinematics from Keplerian rotation, identification of a disk requires to distinguish the disk from those structures in terms of kinematics \citep[e.g.,][]{ohas14}. The outflow can be distinguished relatively easily because its morphology is elongated in the direction perpendicular to the associated disk inclined from the line of sight \citep[e.g.,][]{Arce:2013aa}. The velocity gradient of an outflow is also expected in a direction perpendicular to that produced by disk rotation \citep[][]{Arce:2013aa}. In contrast to outflows, envelopes are hard to distinguish from disks because envelopes often show a flattened, disk-like morphology and are rotating like disks \citep{Galli:1993aa, Galli:1993ab}. The rotation in an envelope is, however, not fast enough to support the material against the gravity of the central star. The material is then infalling in an envelope toward the associated disk. The infalling material is thought to conserve its specific angular momentum ($j$), producing the rotational velocity inversely proportional to the distance from the central protostar \citep{Ulrich:1976aa}:
\begin{eqnarray}    
    v_\mathrm{rot}(r) = \frac{j}{r}.
\end{eqnarray}
This rotational velocity has a steeper radial profile ($\vrot \propto r^{-1}$) than Keplerian disk ($\vrot \propto r^{-0.5}$) and thus cannot support the material against the gravity. It is, therefore, necessary to verify whether the rotation around a protostar is Keplerian rotation or not through observations in order to identify a protostellar disk. Protostellar disks have recently gained a spotlight as the place for planet formation because more evolved, Class $\II$ disks may not be massive enough to form Jupiter-mass planets \citep[][]{Manara:2018aa}. Recent observational works also reported several Class 0/I disks with substructures \citep[e.g.,][]{ALMA-Partnership:2015aa, Sheehan:2018aa, Sheehan:2020aa, Segura-Cox:2020aa, Yamato:2023aa}, which are often thought to be a sign of planet formation. The identification of disks is essential to the study of planet formation during the protostellar phase.

\begin{figure*}[thbp]
\centering
\includegraphics[width=1\textwidth]{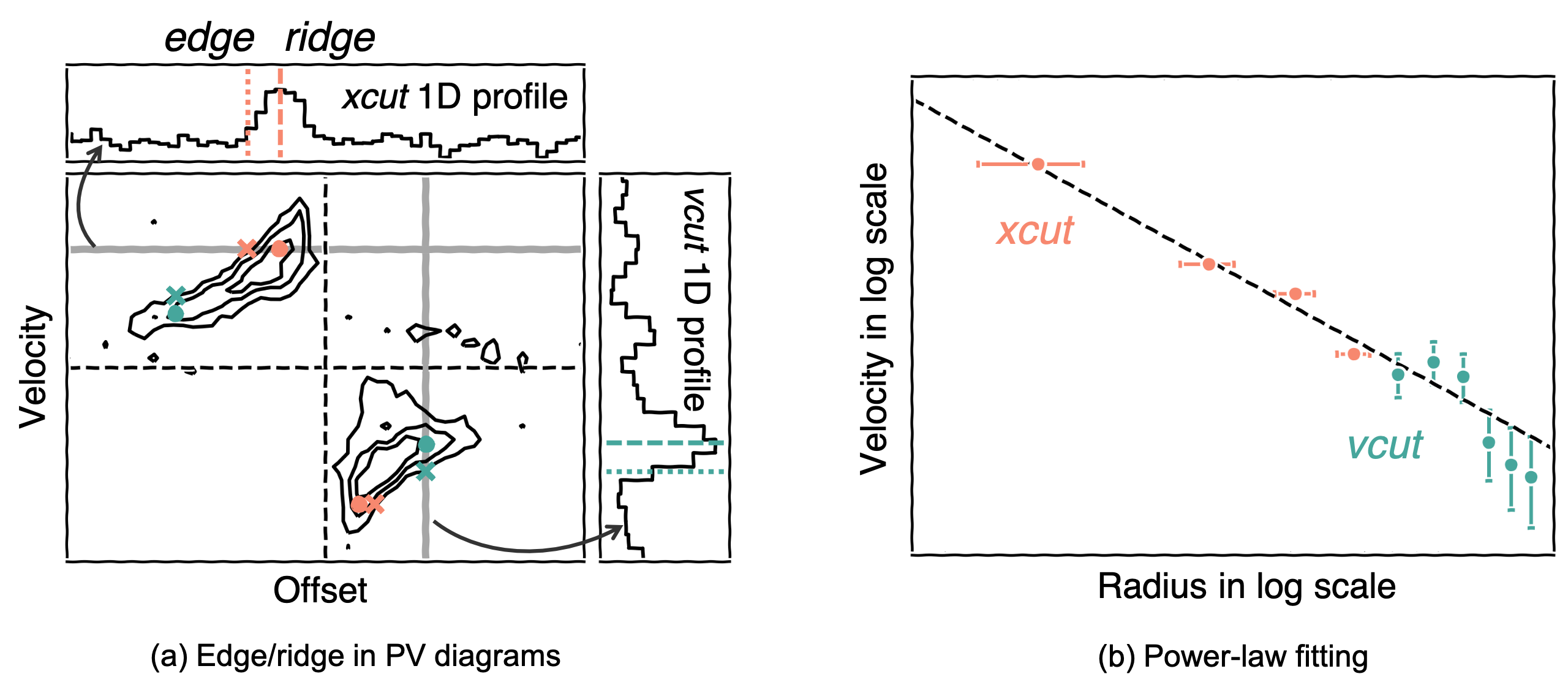}
\caption{Two steps of \texttt{pvanalysis}. (a) Edge/ridge radii and velocities are extracted from the PV diagram, respectively, and then (b) the obtained edge/ridge points are fitted with a power-law function. The left panel illustrates a PV diagram and edge/ridge points derived from \textit{xcut} and \textit{vcut} 1D profiles of the PV diagram. Pink and green markers indicate xcut and vcut data points, respectively. Crosses and circles in the PV diagram denote edge and ridge points, respectively. Dotted and dashed lines in the xcut and vcut 1D profiles also indicate edge and ridge points, respectively. The right panel shows a figure of the fitting of a power-law function to edge/ridge points.
\label{fig:summary_pvana}}
\end{figure*}
Previous works have suggested two directions for identifying protostellar disks. One direction is to construct models, including a disk, and compare them with observational data. Recent sophisticated models of a protostellar system incorporate various calculations, which provide quantities such as a temperature distribution based on a given stellar luminosity and a given density field, non-LTE population of each quantum energy level of a given molecule, the intensity at each position on the plane of the sky through radiative transfer \citep[e.g.,][]{Sheehan:2017aa}. Finding a model that reproduces a given observational data set the best can cost more computational time as more realistic conditions and more free parameters are considered. It is no longer very rare for observational studies to include parameter search that takes a month or longer time scale.
The other direction is to analyze observational data in detail focusing on observable features expected in specific physical structures, such as a disk.
Such analytic estimates have advantages of the lower computational cost and simpler processes than model fitting including radiative transfer with a number of free parameters. Analytic methods also enable us to focus on a specific quantity, such as the central stellar mass, without complicated parameter degeneracy. The python package of \texttt{pvanalysis} implemented in the \texttt{SLAM} (Spectral Line Analysis/Modeling) code \citep{as.sa23}\footnote{This tool is public on GitHub: \url{https://github.com/jinshisai/SLAM}.} has been developed to identify protostellar disks embedded in a protostellar envelope. In this article, we summarize its methodologies and a basic usage, and introduce an application to observational data of this tool.

\section{Methodologies \label{sec:methodologies}}
The \texttt{pvanalysis} tool performs a two-steps analysis using a given position-velocity (PV) diagram made by cutting a 3D cube (2D space projected on the plane of the sky plus 1D line-of-sight velocity) obtained from emission line observations along the major axis of the expected disk. Since rotation causes a velocity gradient along the major axis, PV diagrams along the major axis show the emission distribution as a function of the rotational velocity $v$ and the offset or radius from the central protostellar position $r$.
The first step of this tool is to extract data points from the PV diagram that represent the relation between the rotational velocity and the radius. The second step is to fit these data points with a power-law function. These processes are visualized in Figure \ref{fig:summary_pvana}. We describe details of each step in the following subsections.

\subsection{Edge and Ridge} \label{sec:edgeridge}
The first step, extracting data points $(r_i, v_i)$ from a PV diagram, is performed through two independent methods. One method uses the emission ridge in 1D profiles along the positional axis (the velocity axis) at a given velocity (position). This ridge method is demonstrated in previous observational studies \citep{ohas14, aso17, yen17, sai20} with discussion about its potential problems \citep{yen13, aso15, sai20}, as well as in a study using synthetic observations of a magnetohydrostatic (MHD) simulation of protostellar evolution \citep{as.ma20}. The other method uses the emission outer edge in the 1D profiles. This edge method was developed using an MHD simulation of protostellar evolution \citep{seif16} and is used in previous observational studies as well \citep[e.g.,][]{alve17}. The advantage and disadvantage of each method are controversial. For example, \citet{mare20} reported that, depending on the spatial resolution, the ridge method can underestimate the central stellar mass by $\sim 30\%$, while the edge method can overestimate it by a factor of $\sim 2$.

The edge and ridge radii are obtained in the 1D profile along the positional axis at the given velocity (\textit{xcut}); those radii are obtained at every channel (Figure \ref{fig:summary_pvana}). The edge radius is defined as the outermost position with a given threshold level (e.g., $5\sigma$) of emission. The uncertainty of the edge radius is calculated as the absolute value of the intensity gradient at the edge radius multiplied by the noise level of the PV diagram. The intensity gradient is calculated by interpolating the 1D profile.
The ridge radius is defined as the intensity-weighted mean position in the 1D profile, or the center derived by Gaussian fitting. The uncertainty of the ridge radius is calculated from the noise level of the PV diagram through the error propagation in the case of the mean position, while it is provided by the Gaussian fitting in the case of the Gaussian center. 
Similarly the edge and ridge velocities are obtained in the 1D profile along the velocity axis at a given position (\textit{vcut}; Figure \ref{fig:summary_pvana}). In the vcut case, those velocities are obtained with a positional separation of half of the beam major-axis, rather than one pixel, to avoid sampling more than the Nyquist rate. The oversampling causes to put unnecessarily high weights on the vcut points, relative to the weights on xcut points, in the $\chi ^2$ fitting with a power-law function.

\begin{figure}[htbp]
\includegraphics[width=0.5\textwidth]{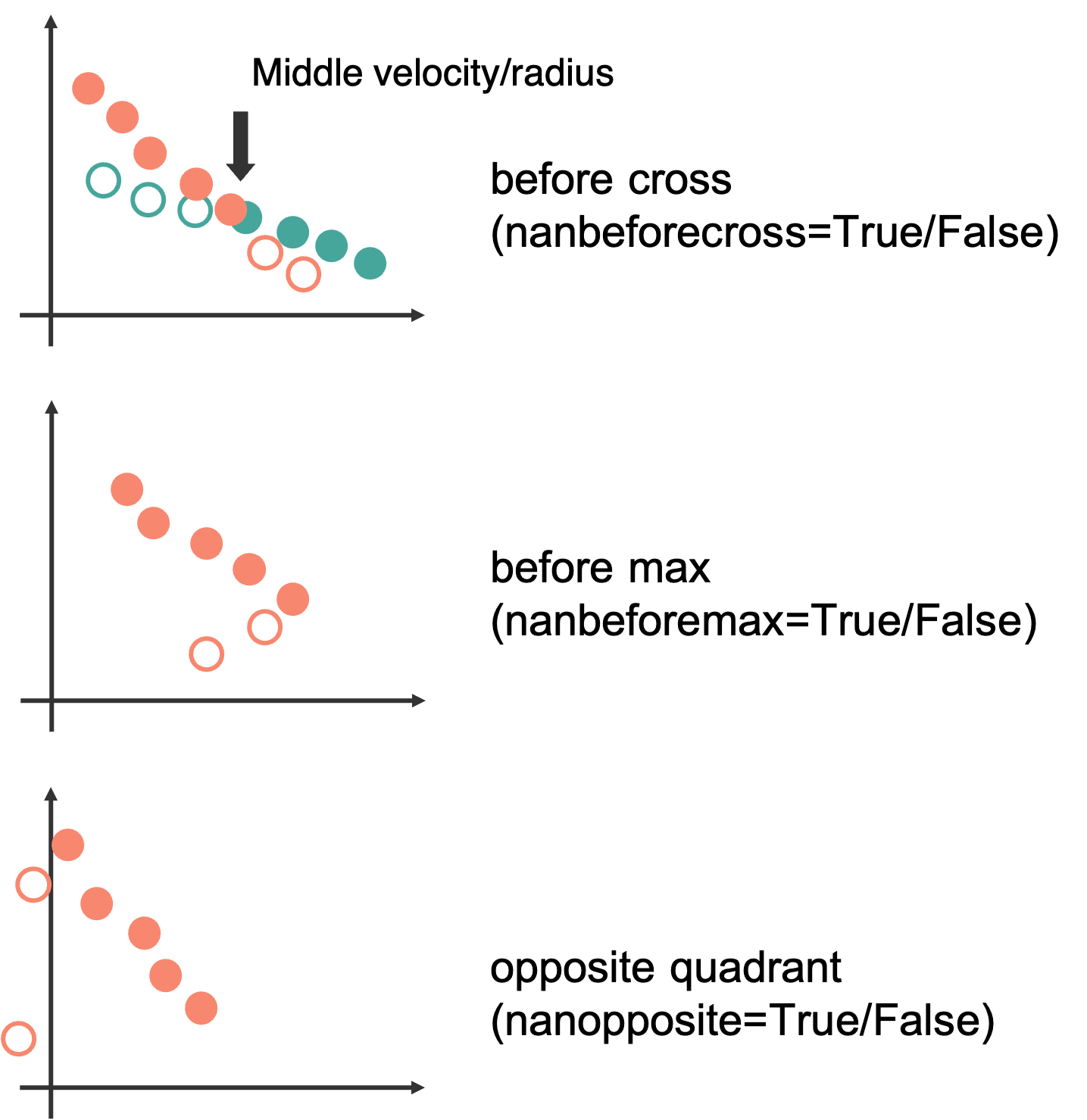}
\caption{Filtering conditions so that edge/ridge points better represent the rotation curve. Filled and open circles indicate data points that will be used for and removed from the rotation curve fitting, respectively. The nanbeforecross parameter removes the vcut-points at smaller radii and the xcut-points at lower velocities than the closest pair to obtain a one-to-one relation between the radius and the velocity. The nanbeforemax parameter removes the vcut(xcut)-points at the radii smaller (velocities lower) than the radius (velocity) of the maximum velocity (radius) to achieve the spin-up relation between the radius and the velocity. The nanopposite parameter removes the points on the unexpected quadrant. These three filters are described in Section \ref{sec:edgeridge} in more detail.
\label{fig:nan_thr}}
\end{figure}

Then, obtained ridge and edge points are passed to filters so that data points better representing the rotation curve feature can be fitted with a power-law function in a latter fitting stage. The filtering conditions are illustrated in Figure \ref{fig:nan_thr}. In each of edge and ridge method, xcut points at velocities higher than a ``middle" velocity and vcut points at radii outer than a ``middle" radius are combined as the final pairs of $(r_i, v_i)$ (top of Figure \ref{fig:nan_thr}). The middle velocity and radius are determined from the closest pair of points between the xcut and vcut, as explained in \citet{as.ma20} in more detail. In addition, in xcut (vcut), the edge/ridge radii (velocities) can be removed if the radii (velocities) are at lower velocities (smaller radii) than the velocity (radius) with the maximum edge/ridge radius (velocity) (middle of Figure \ref{fig:nan_thr}). Furthermore, the edge/ridge radii or velocities can be removed if those are located in the opposite quadrant (bottom of Figure \ref{fig:nan_thr}). Removing points is done separately between the redshifted and blueshifted velocities (or positive and negative offsets).

\subsection{Power-law Fitting} \label{sec:plfitting}
The pairs of $(r_i, v_i)$ obtained in the edge and ridge methods are separately fitted with a single- or double-power function as follows:
\begin{eqnarray}
v = {\rm sgn}(r)\ v_\mathrm{b} \left(\frac{|r|}{r_\mathrm{b}}\right) ^{-p}+v_{\rm sys}, \nonumber\\
\label{eq:power}
p = p_{\rm in}\ \ {\rm if}\ \ |r|<r_\mathrm{b}\ \ {\rm else}\ \ p_{\rm in} + dp,
\end{eqnarray}
where $v_\mathrm{b}$ is the velocity at the radius of $r_\mathrm{b}$, $p_{\rm in}$ is the power-law index, $dp$ is the difference of the index between the inner and outer parts, and $v_{\rm sys}$ is the systemic velocity. The sign of the first term on the left hand side of Equation (\ref{eq:power}) means that emission tracing rotation appears in the first and third quadrants in the PV diagram. When the rotation appears in the second and forth quadrants, this sign is inverted. $dp$ is limited to be $\geq 0$ to produce a steeper or the same index outside $r_\mathrm{b}$, such as the inner Keplerian rotation versus the outer rotation where the angular momentum is conserved. $dp=0$ means the single power-law fitting. Equation (\ref{eq:power}) provides a function of radius, defined here as $V_{\rm fit}(r)$. When $dp \geq 0$, the inverse function of velocity $R_{\rm fit}(v)$ can also be defined. Using these functions, $\chi ^2$ to be minimized is defined as 
\begin{eqnarray}
\chi ^2 &=& \sum _{i~\rm for~xcut} \left(r_{i, {\rm obs}}- R_{\rm fit}(v_{i, {\rm obs}})\right) / \sigma^{2}_{i, {\rm obs}} \nonumber \\
&&+ \sum _{j~\rm for~vcut} \left(v_{j, {\rm obs}} - V_{\rm fit}(r_{j, {\rm obs}})\right)^2 / \sigma^{2}_{j, {\rm obs}},
\label{eq:chi2}
\end{eqnarray}
where $\sigma$ is the uncertainty of each radius/velocity defined in Section \ref{sec:edgeridge}: The edge uncertainty is calculated through the emission gradient, the mean-ridge uncertainty is calculated through the error propagation, and the Gaussian-ridge uncertainty is calculated through the 1D Gaussian fitting. If the vcut points are sampled on every pixel, the sum of vcut becomes more dominant in Equation (\ref{eq:chi2}) with smaller pixels by summing multiple points within the half-beam, which have effectively the same information. Then, the fitting process will search for a solution that only minimizes the sum of vcut, ignoring the sum of xcut. To avoid such imbalance, the vcut-points are sampled at the half-beam interval (Section \ref{sec:edgeridge}).

The $\chi ^2$ is minimized by the Markov chain Monte Carlo method using the open python package {\tt emcee} \citep[][]{fore13} widely used in modelings of YSOs. We adopt this method because it searches for parameters efficiently around the best parameters, and it shows acceptable parameter ranges and degeneracy among parameters through the posterior distribution.
By default, the numbers of walkers per parameter, burn-in steps, and steps after the burn-in are 16, 2000, and 2000, respectively. The 50 percentile of each parameter is adopted as the best-fit value. The uncertainty of each parameter is calculated as the 16 and 84 percentiles. These numbers will be justified in the actual application to observational data in Section \ref{sec:act}.
The search ranges for $r_\mathrm{b}$ and $v_\mathrm{b}$ are limited between the maximum and minimum of the edge/ridge radii and velocities, respectively. The ranges for $p_{\rm in}$ and $dp$ are limited to $[0.01, 10]$ and $[0, 10]$, respectively. 
These ranges are wide enough because the indices expected from physics are around 0.5 (Keplerian rotation) or 1.0 (conservation of angular momentum).
The range for $v_{\rm sys}$ is limited to $[-1~\kms, 1~\kms]$ relative to a given systemic velocity.
This range is much wider than typical velocity resolutions for YSO studies, and thus vsys can be estimated within this range before using our tool.
No other prior is set for the MCMC fitting. In the case of single power-law fitting, $v_\mathrm{b}$ is fixed at the logarithmic center of the search range of $v_\mathrm{b}$: When the range is $[v_{\rm min}, v_{\rm max}]$, the logarithmic center is $\sqrt{v_{\rm min}v_{\rm max}}$.

\section{Usage and Parameters} \label{sec:usage}
In this section, we describe actual usage and parameters of the \texttt{pvanalysis} tool. In addition to the detailed descriptions in this section, an example script can be found in our GitHub repository\footnote{\url{https://github.com/jinshisai/SLAM}}.

\subsection{Inputs}
The each analysis step described in Section \ref{sec:methodologies} is provided as a method for a python class, \texttt{PVAnalysis}, in this tool. The instantiation requires five parameters, including the input data:
\begin{verbatim}
impv = PVAnalysis(fitsfile, rms, vsys, dist, incl)
\end{verbatim}
\begin{description}
    \item[{\tt \bf fitsfile} ---] Path to the input fits file of a PV diagram.
    \item[{\tt rms} ---] The root-mean-square (rms) noise level of the PV diagram.
    \item[{\tt vsys} ---] The systemic velocity of the object in the unit of $\mathrm{km~s}^{-1}$. When the systemic velocity $v_{\rm sys}$ is a free parameter, this value is the center of the range over which the free parameter is searched.
    \item[{\tt dist} ---] The distance to the object in the unit of pc to convert the unit of radius from arcsecond to au.
    \item[{\tt incl} ---] The inclination angle of the disk in the unit of degree, which is used to calculate the central stellar mass. If not given, an edge-on configuration (i.e., an inclination angle of $90^\circ$) is assumed.
\end{description}
The input of observational data must be a fits file of a PV diagram. The input fits file is supposed to be two or three dimensions: it should have offset and velocity/frequency axes, or have a one-dimensional stokes axis (i.e., with only stokes I and no polarization) in addition to the offset and velocity/frequency axes. It is assumed that the offset axis is the first axis, and  the stokes axis is the third axis if it is present. The second axis can be either velocity or frequency. If the second axis is given as velocity, it must be in the unit of $\mathrm{m~s}^{-1}$, and then it is converted in the unit of $\mathrm{km~s}^{-1}$. If the second axis is given as frequency, it is automatically converted to velocity based on the line rest frequency provided in the fits header.

The first phase of the analysis where edge and ridge points are determined (with a method of \texttt{get\_edgeridge}) requires inputs to limit which data of a PV diagram are used. For the simplest use, one can only give the output file name and the \texttt{thr} parameter:
\begin{verbatim}
impv.get_edgeridge(outname=`filename', thr=5)
\end{verbatim}
\begin{description}
    \item[{\tt outname} ---] Header name of output files. The output file will be, for example, outname.ridge.dat.
    \item[{\tt thr} ---] The pixels where the intensity is higher than this threshold are used for determining the ridge points. The outermost radius (the highest velocity) with this value is the edge radius (velocity). The unit is the given rms noise level. Defaults to 5.
\end{description}
Then, edge/ridge radii and velocities will be calculated using the whole radial and velocity ranges of a given PV diagram following the method and criteria described in Section \ref{sec:methodologies}. One might want to calculate edge/ridge points only within limited radial and/or velocity ranges. The following parameters can be used for such a case in the method of \texttt{get\_edgeridge}:
\begin{description}
    \item[{\tt xlim} ---] When ${\rm xlim}=(a, b, c, d)$, the edge and ridge radii are removed unless they are between $a$ and $b$ or $c$ and $d$.
    \item[{\tt vlim} ---] Same as {\tt xlim} but for the velocities.
    \item[{\tt Mlim} ---] When ${\rm Mlim}=(a, b)$, the edge and ridge points $(r_i, v_i)$ are removed unless $r_i V_i^2/G$ is between $a$ and $b$.
    \item[{\tt use\_position} ---] Whether this process uses the edge and ridge radii, i.e., xcut.
    \item[{\tt use\_velocity} ---] Whether this process uses the edge and ridge velocity, i.e., vcut.
\end{description}
These inputs allow to remove points that appear suffering from any effect that this tool does not suppose, such as the resolving-out effect in interferometric observations and a low signal-to-noise ratio (S/N). In addition, some of the filtering processes could be turned-on and off through the following parameters in the method of \texttt{get\_edgeridge}. All of which are True by default.
\begin{description}
    \item[{\tt nanbeforecross} ---] This switches on the removal of lower velocities and smaller radii to combine the xcut and vcut points.
    \item[{\tt nanbeforemax} ---] This switches on the removal of the edge and ridge radii (velocities) at velocities (radii) lower (smaller) than the velocity (radius) with the maximum radius (velocity).
    \item[{\tt nanopposite} ---] This switches on the removal of the points on the unexpected quadrants.
\end{description}
To avoid that only few points that has S/N much higher than others control the power-law function fitting, the inputs of {\tt minabserr} and {\tt minrelerr} are introduced in the method of \texttt{get\_edgeridge}:
\begin{description}
    \item[{\tt minabserr} ---] This sets the minimum absolute error for each edge or ridge point. The unit is the major-axis of the observational beam for the radius, while it is the channel width for the velocity. When the absolute error bar for an edge or ridge point is smaller than this value, it is replaced with this value. Defaults to 0.1.
    \item[{\tt minrelerr} ---] This sets the minimum relative error for each edge or ridge point. When the relative error bar for an edge or ridge point is smaller than this value, it is replaced with this value. Defaults to 0.01.
\end{description}
Ridge points can be derived through not only the intensity-weighted mean but also the Gaussian fitting. They can be switched using the following parameter in the method of \texttt{get\_edgeridge}:
\begin{description}
    \item[{\tt ridgemode} ---] `mean' or `gauss'. The method for determining the ridge points: the intensity-weighted mean radius (velocity) or the Gaussian-center radius (velocity). Defaults to `mean'.
\end{description}
\vskip\baselineskip

The power-law fitting part (\texttt{fit\_edgeridge}) requires inputs to define the model function.
\begin{description}
    \item[{\tt include\_vsys} ---] Whether the systemic velocity is a free parameter.
    \item[{\tt include\_dp} ---] Whether the difference between the inner and outer power-law indices is a free parameter.
    \item[{\tt include\_pin} ---] Whether the inner power-law index is a free parameter.
    \item[{\tt fixed\_pin} ---] The inner power-law index is fixed to this value when it is not a free parameter.
    \item[{\tt fixed\_dp} ---] The difference between the inner and outer power-law indices is fixed to this value when it is not a free parameter.
\end{description}
For example, the following command means that the model function is the Keplerian rotation, which has only one free parameter:
\begin{verbatim}
impv.fit_edgeridge(include_vsys=False,
                   include_dp=False,
                   include_pin=False,
                   fixed_pin=0.5,
                   fixed_dp=0,
                   calc_evidence=False).
\end{verbatim}
The last input, \texttt{calc\_evidence}, is an optional parameter and allows you to calculate a model evidence using the \texttt{dynesty} package \citep{Speagle:2020aa} when it is set to \texttt{True}. A model evidence is the integral of a likelihood function over the parameter space. One can calculate the Bayes factor by taking a evidence ratio between two models, which would be useful for discussing the goodness of fits among different models (e.g., \texttt{include\_pin=True} versus \texttt{False} in Section \ref{sec:act}).

In addition, the \texttt{PVAnalysis} class provides methods for writing the obtained edge and ridge points to a text file (\texttt{write\_edgeridge(outname='filename')}), outputting the fitting results in the terminal window (\texttt{output\_fitresult()}), and plotting PV diagrams with the ridge/edge points and the best-fit curves (\texttt{plot\_fitresult()}).

\subsection{Outputs}

\begin{figure}[htbp]
\centering
\includegraphics[width=0.8
\columnwidth]{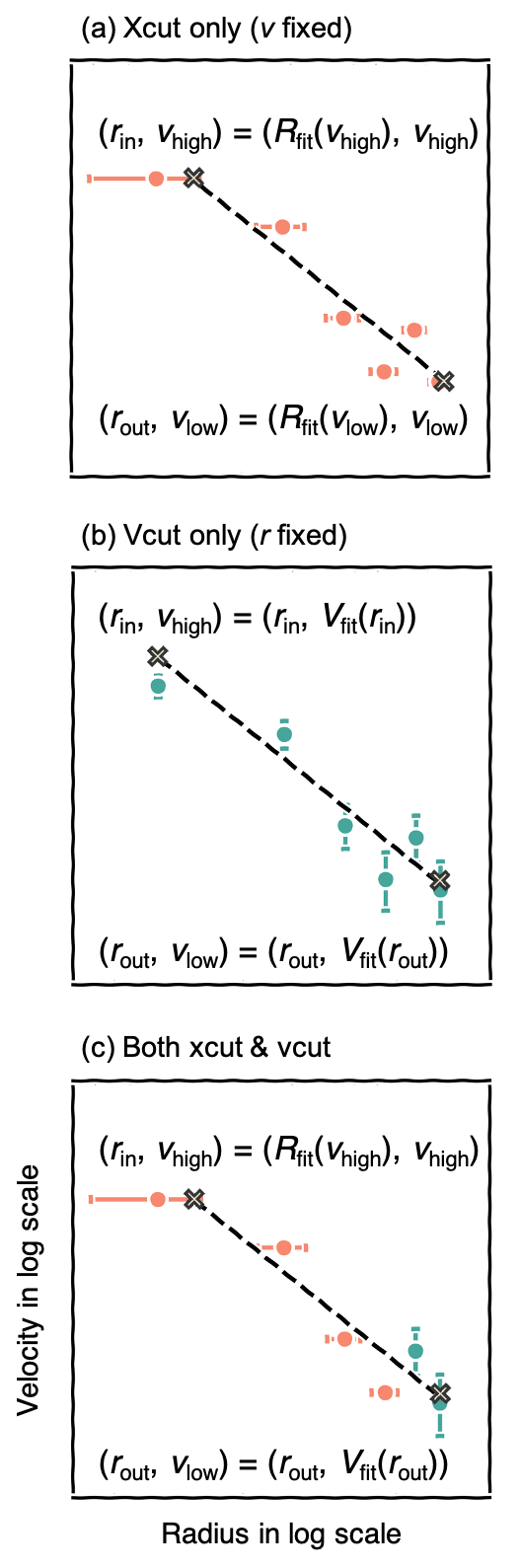}
\caption{Illustration of definitions of ($r_\mathrm{in}$, $v_\mathrm{high}$) and ($r_\mathrm{out}$, $v_\mathrm{low}$) in cases of (a) xcut-only, (b) vcut-only, and (c) the combination of xcut and vcut. Dashed lines indicate the best-fitted power-law function and crosses denote points where ($r_\mathrm{in}$, $v_\mathrm{high}$) and ($r_\mathrm{out}$, $v_\mathrm{low}$) are defined.
\label{fig:inout-def}}
\end{figure}

The main output is the best-fit parameters, $r_\mathrm{b}$, $v_\mathrm{b}$, $p_{\rm in}$, $dp$, and $v_{\rm sys}$ with their error bars; the error bar is zero when the parameter is not free. In addition, ranges of radius and velocity are calculated using the best-fit model function, as summarized in Figure \ref{fig:inout-def}. With only xcut, the highest velocity, $v_{\rm high}$, is the highest velocity among the velocities where edge/ridge radii are obtained. Then, the innermost radius $r_{\rm in}$ is calculated from Equation (\ref{eq:power}) as the radius at $v=v_{\rm high}$, i.e., $r_{\rm in}=R_{\rm fit}(v_{\rm high})$. The pair of $(r_{\rm out}, v_{\rm low})$ is similarly defined in the case with only xcut. 
With only vcut, the innermost radius, $r_{\rm in}$, is the smallest radius among the radii where edge/ridge velocities are obtained. Then, the highest velocity $v_{\rm high}$ is calculated from Equation (\ref{eq:power}) as the velocity at $r=r_{\rm in}$, i.e., $v_{\rm high}=V_{\rm fit}(r_{\rm in})$. The pair of $(r_{\rm out}, v_{\rm low})$ is similarly defined.
With the combination of xcut and vcut, $(r_{\rm in}, v_{\rm high})$ is defined as in the case with only xcut, while $(r_{\rm out}, v_{\rm low})$ is defined as in the case with only vcut. This is inner, higher-velocity points in the xcut set and outer, lower-velocity points in the vcut set are supposed to be combined to obtain the final edge/ridge points (Section \ref{sec:edgeridge}) in the case with both xcut and vcut. These definitions of $(r_{\rm in}, v_{\rm high})$ and $(r_{\rm out}, v_{\rm low})$ for different cases are illustrated in Figure \ref{fig:inout-def}.

From the pair of $(r_\mathrm{b}, v_\mathrm{b})$, the central stellar mass can be calculated as $M_* = v_\mathrm{b} ^2 r_\mathrm{b} / G / \sin ^2 i$, where $G$ is the gravitational constant and $i$ is the inclination angle. Estimation of the inclination angle is out of the scope of this tool; the inclination angle is often estimated from the aspect ratio of the associated (sub)mm continuum emission or independent modeling of the associated outflow. If the power-law index $p_{\rm in}$ is significantly different from the Keplerian index 0.5, the estimated $M_*$ does not have a physical meaning of the central stellar mass. This tool also provides two more masses: $M_{\rm in}=v_{\rm high} ^2 r_\mathrm{in} / G / \sin ^2 i$ and $M_{\rm out}=v_{\rm low} ^2 r_{\rm out} / G / \sin ^2 i$.

This tool also outputs two text files and four figure files. Two text files are produced by the method of \texttt{write\_edgeridge()}. One of the two text files shows the pairs of ridge $(r_i, v_i)$ with their error bars. The error bar of $r_i$ ($v_i$) is zero for the vcut (xcut) points. The other text file similarly shows the edge pairs.
Two figures show the corner plots produced by the MCMC fitting, which are produced by the method of \texttt{fit\_edgeridge()}. One figure is for the ridge fitting, and the other is for the edge fitting. The remaining two figure are produced by the method of \texttt{plot\_fitresult()} and shows a PV diagram with the edge and ridge points and the best-fit curves; one figure for the linear scale and one figure for the logarithmic scale. The logarithmic PV diagram is produced by calculating the mean between the first- and third-quadrants (or second- and forth-quadrants) of the linear PV diagram. When the systemic velocity is a free parameter, the mean of the edge- and ridge-systemic velocities is used to calculate the logarithmic PV diagram.

\section{Examples with Observational Data} \label{sec:act}
\begin{figure}[htbp]
\begin{tabular}{c}
\begin{minipage}[t]{0.5\textwidth}
\centering
\includegraphics[width=1\textwidth]{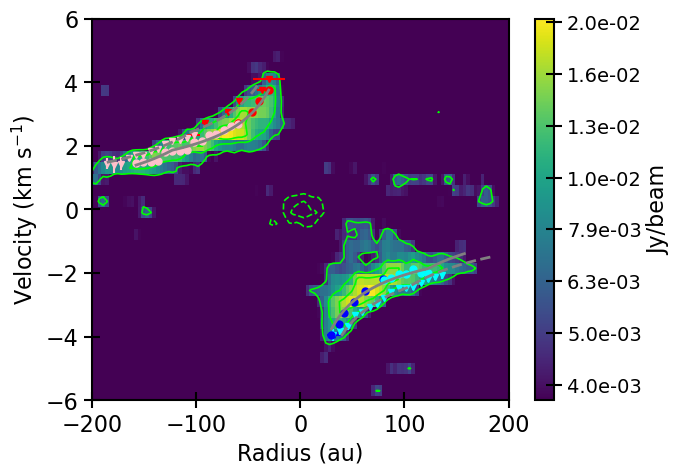}
\subcaption{}
\end{minipage}
\\
\begin{minipage}[t]{0.5\textwidth}
\centering
\includegraphics[width=1\textwidth]{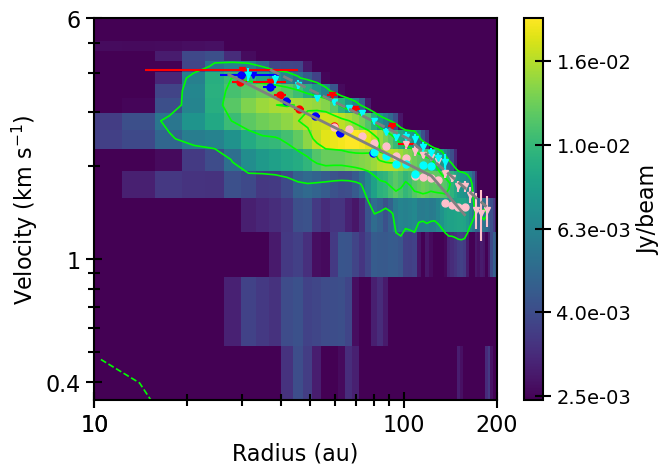}
\subcaption{}
\end{minipage}
\end{tabular}
\caption{Edge points, ridge points, and the best-fit power-law function overlaid in a position-velocity diagram on the (a) linear and (b) logarithmic scales. The velocity is relative to the systemic velocity $6.4~\kms$. The triangles and circles are the edge and ridge points, respectively. The red and blue points are the xcut points, while the pink and cyan points are vcut points. The xcut points have error bars in the positional direction. The vcut points have error bars in the velocity direction. The dashed and solid curves show the best-fit power-law functions for the edge and ridge points, respectively. The contour levels are $\pm 3$, $\pm 6$, and $\pm 9\sigma$, where $1\sigma$ is $1.7~\mJB$.
\label{fig:linlog}}
\end{figure}

Figure \ref{fig:linlog} shows the result of our tool using an actual observational data set\footnote{The project codes of the data used in this demonstration are 2013.1.01086.S and 2015.1.01415.S. The targeted protostar has a name of TMC-1A. The angular resolution is $0\farcs 12\times 0\farcs 08$ (P.A.=$35\arcdeg$). The velocity resolution is $0.35~\kms$. The PV diagram is produced with a cut at P.A.=$75\arcdeg$.} of a molecular line in a protostellar system that consists of a disk and an envelope. Figure \ref{fig:linlog}(a) shows a PV diagram along the major axis of the disk. The abscissa is the distance from the central protostellar position, i.e., radius in the unit of au (distance is 140~pc), while the ordinate is the velocity relative to the systemic velocity ($6.4~\kms$). This diagram shows a tendency that the emission is closer to the center at higher velocities. Because PV diagrams along the disk major axis are used to investigate the relation between the rotational velocity and the radius, this tendency can be interpreted as a spin-up rotation {\rm i.e., the rotational velocity is higher at a position closer to the central protostellar position}. The points in Figure \ref{fig:linlog} are the edge/ridge xcut/vcut points; the triangles and circles are the edge and ridge points, respectively, while the red/blue and pink/cyan points denote the xcut and vcut points, respectively. These points successfully trace the spin-up tendency in the PV diagram.

\begin{figure*}[htbp]
\centering
\includegraphics[width=0.85\textwidth]{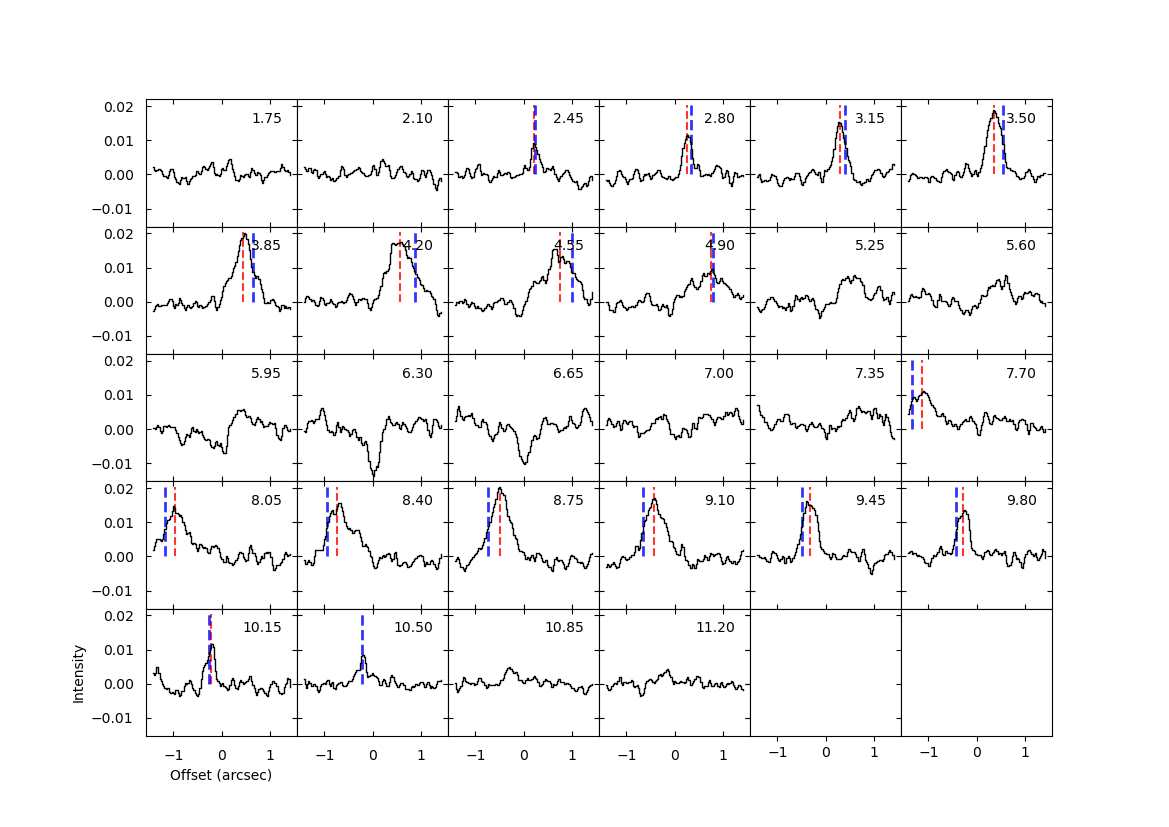}
\caption{The xcut points plotted in 1D intensity profiles. The blue and red points are the edge and ridge points, respectively. Each panel shows a 1D intensity profile where the abscissa is the offset from the central position. The velocity (not relative to the systemic velocity but absolute) of each profile is denoted on the upper right corner of each panel in the unit of $\kms$.
\label{fig:x1d}}
\end{figure*}

\begin{figure*}[htbp]
\centering
\includegraphics[width=0.85\textwidth]{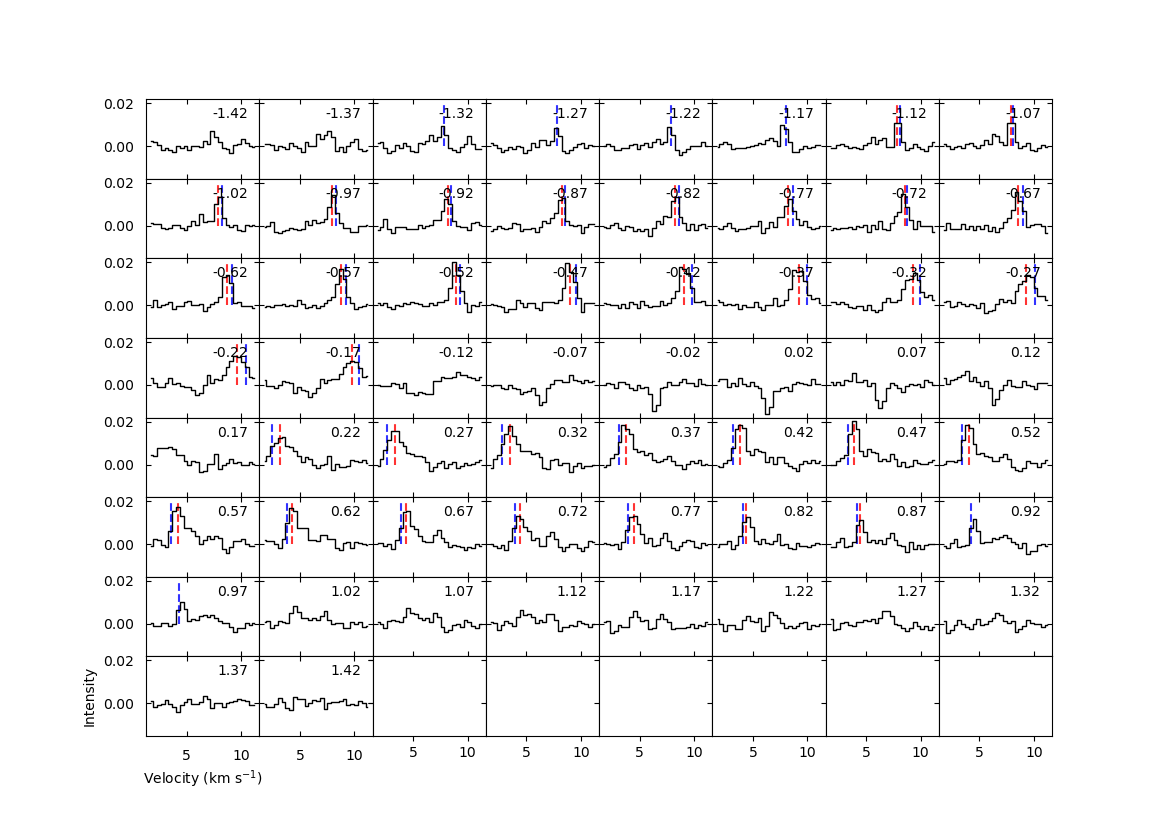}
\caption{The vcut points plotted in 1D intensity profiles. The blue and red points are the edge and ridge points, respectively. Each panel shows a 1D intensity profile where the abscissa is the velocity (not relative to the systemic velocity but absolute). The positional offset of each profile is denoted on the upper right corner of each panel in the unit of arcsecond.
\label{fig:v1d}}
\end{figure*}

Figures \ref{fig:x1d} and \ref{fig:v1d} show the edge and ridge points plotted in 1D intensity profiles with which the edge or ridge points are obtained. The xcut and vcut points in these figures are before combined. The panels without the edge or ridge point (blue or red dashed line) have S/N ratios not enough to obtain these points. The offsets of the vcut profiles are sampled at the Nyquist rate. The ridge points trace the peak offset and velocity of each 1D intensity profile, and the edge points trace an outer position and velocity of the profile, which demonstrate that the edge and ridge points are well-defined. 


\clearpage

To describe the radial profile of rotational velocity, a broken power-law function is adopted that fixes $p_{\rm in}=0.5$ (Keplerian rotation) and $v_{\rm sys}=6.4~\kms$, while varying $v_\mathrm{b}$, $r_\mathrm{b}$, and $dp$ in the MCMC fitting. The gray curves in Figure \ref{fig:linlog} show the best-fit power-law functions; the dashed and solid curves show those for the edge and ridge points, respectively. 

\begin{figure}[htbp]
\begin{tabular}{c}
\begin{minipage}[t]{0.5\textwidth}
\centering
\includegraphics[width=1\textwidth]{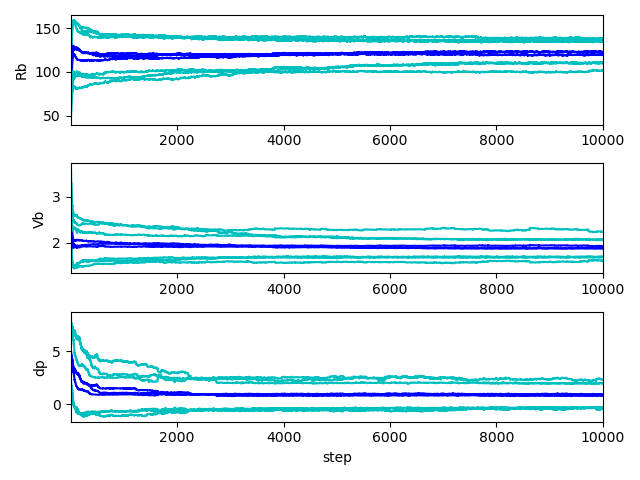}
\subcaption{parameters}
\end{minipage}
\\
\begin{minipage}[t]{0.5\textwidth}
\centering
\includegraphics[width=1\textwidth]{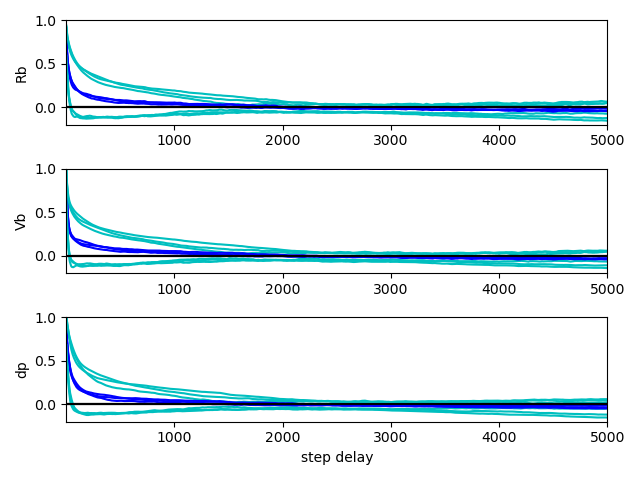}
\subcaption{autocorrelations}
\end{minipage}
\end{tabular}
\caption{The three parameter (panel a) and their autocorrelation (panel b) obtained in the MCMC fitting. The denser blue lines show the mean, while the cyan lines show the mean $\pm$ the standard deviation among 16 chains. Different lines with the same color show three independent results to inspect the impact of any random number in the MCMC fitting. The autocorrelation is calculated using neighboring 5000 steps and normalized by the maximum value at the no delay point.
\label{fig:converge}}
\end{figure}

Figure \ref{fig:converge} shows the mean (panel a) and the autocorrelation (panel b) of the three parameters sampled in the MCMC fitting as a function of the step. This figure shows the results for the ridge points; the results for the edge points similarly show the following results. In Figure \ref{fig:converge}(a), the denser blue lines show the mean values among the 16 chains, while the cyan lines show the mean $\pm$ the standard deviation among the chains. Different lines with the same color show three independent results to inspect the effect of any random number in the MCMC fitting. In Figure \ref{fig:converge}(b), the denser blue lines show the autocorrelation of the parameter offset from the mean among the steps. This autocorrelation is calculated using neighboring 5000 steps in Figure \ref{fig:converge}(a) and normalized by the maximum value at the no delay point. The cyan lines show the mean $\pm$ the standard deviation of the autocorrelation among the chains. The meaning of different lines with the same color is the same as in Figure \ref{fig:converge}(a). These figures show that the sampled parameters are converged after 2000 steps, with the adopted number of walkers per parameter 16. In other words, the parameters sampled after 2000 steps are not correlated with the initial values. This supports the numbers of walkers and the burn-in steps (Section \ref{sec:plfitting}).

\begin{figure}[htbp]
\begin{tabular}{c}
\begin{minipage}[t]{0.5\textwidth}
\centering
\includegraphics[width=1\textwidth]{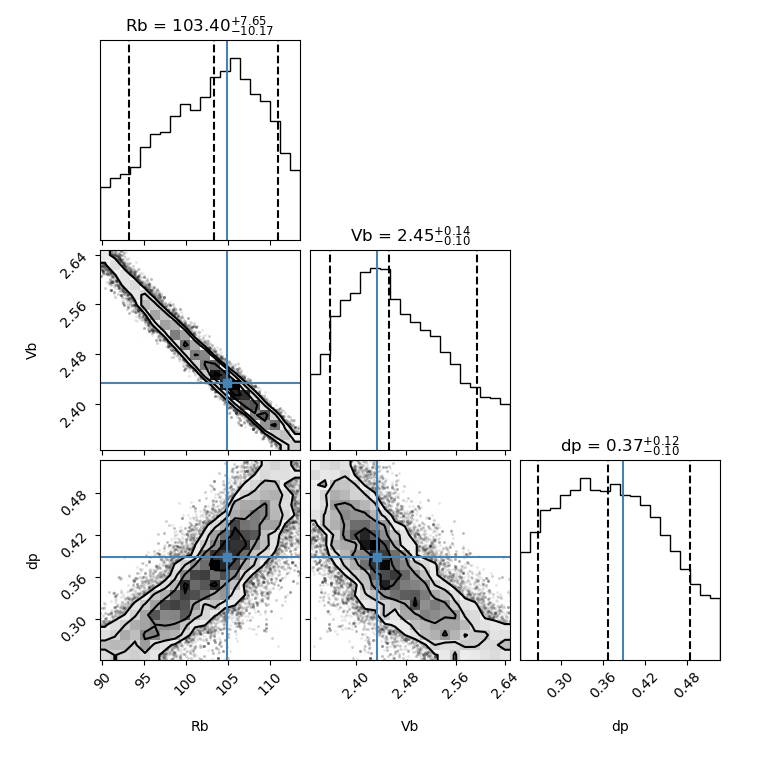}
\subcaption{}
\end{minipage}
\\
\begin{minipage}[t]{0.5\textwidth}
\centering
\includegraphics[width=1\textwidth]{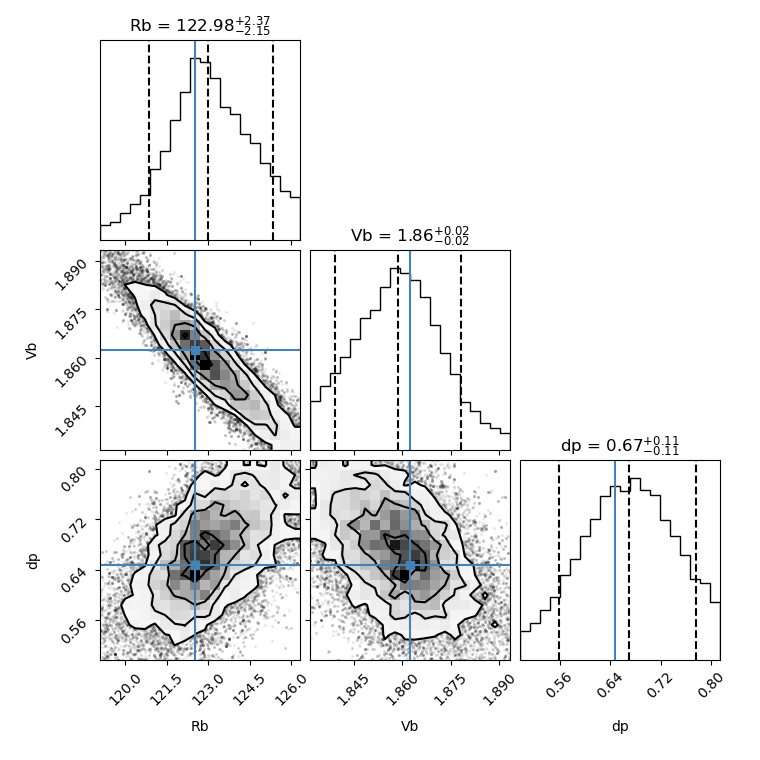}
\subcaption{}
\end{minipage}
\end{tabular}
\caption{Corner plots of the three parameters, $r_\mathrm{b}$ in the unit of au, $v_\mathrm{b}$ in the unit of $\kms$, and $dp$, searched in the MCMC fitting with {\tt emcee} to the (a) edge and (b) ridge points. The blue lines and points show the parameter set that provides the smallest $\chi ^2$. The dashed lines show the 16, 50, and 84 percentiles of the posterior distribution of each parameter.
\label{fig:corner}}
\end{figure}

Figures \ref{fig:corner} shows corner plots of the posterior distribution of the three free parameters obtained in the MCMC fitting, using 2000 steps after the burn-in. The posterior distribution is well sampled with the 2000 steps. The blue lines and points denote the parameter set that provides the smallest $\chi ^2$. This parameter set is close to the 50 percentile parameter set, indicating that the 50 percentile can be adopted as the best-fit parameter set. The best-fit three parameters are listed in Table \ref{tab:best}. The evidence of the posterior distribution was calculated through \texttt{dynesty} to be $\sim 2.3\times 10^{-15}$ for the edge points and $\sim 1.1\times 10^{-41}$ for the ridge points. The parameter $dp$ means that the outer power-law index is $\sim 0.9$ in the edge method and $\sim 1.2$ in the ridge method, which may suggest a rotation with a constant specific angular momentum ($\vrot\propto r^{-1}$). Because the inner power-law index is assumed to be the Keplerian-rotation index, $r_\mathrm{b}$ is interpreted as the boundary radius between a Keplerian disk and a surrounding envelope: $r_\mathrm{b}\sim 100$--$120$~au. This radius is consistent with the disk radius reported by \citet{aso15} with multiple analyses and modeling to the same target. The difference of $r_\mathrm{b}$ between the edge and ridge methods is $>2$ times larger than the error bar of each $r_\mathrm{b}$. Similarly, the difference of $v_\mathrm{b}$ between the edge and ridge methods is much larger than the error bar of each $v_\mathrm{b}$.

Using the best-fit power-law function, the two masses $M_{\rm in}$ and $M_{\rm out}$ are calculated; their error bars are also calculated through error propagation (Table \ref{tab:cal}). Previous works estimated the inclination angle of this system to be $i=51\arcdeg$ \citep{hars18} and $53\arcdeg$ \citep{aso21}. Hence, we adopt $i=50\arcdeg$ for this example calculation. $M_b$ is the same as $M_{\rm in}$ in this fitting because the inner power-law index $p_{\rm in}$ is fixed at 0.5 (Keplerian rotation). Hence, $M_{\rm in} (=M_b)$ is the central stellar mass estimated from this fitting. The edge mass and ridge mass are thought to provide an upper limit and a lower limit of the central stellar mass, respectively, i.e., $M_*\sim $0.8$-$1.2$~\Msun$. The mass before the inclination correction, $M_* \sin ^2 i=0.5-0.7$, is consistent with the one reported for the same target by \citet{aso15}, $M_* \sin ^2 i=0.56\pm 0.05~\Msun$. The derived error bars for $M_{\rm in}$ and $M_{\rm out}$ are a few times smaller than the difference between the edge and ridge masses. This suggests that the uncertainty of the stellar mass estimation is dominated by the difference between the edge and ridge methods. Because $dp$ is significantly different from zero, $M_{\rm out}$ no longer has a meaning of the central stellar mass. The best-fit power-law function is also used to calculate the $r$ and $v$ ranges of the edge and ridge points (Table \ref{tab:cal}). Using this range, $M_{\rm out}$ can be converted to a specific angular momentum: $j=\sqrt{GM_{\rm out}r_{\rm out}}\sim $360 and 290 $\kms$~au in the edge and ridge methods, respectively.

\begin{table}[t!]
\caption{The best-fit parameters in Section \ref{sec:act}}
\centering
\begin{tabular}{ccccc}
\toprule
& $r_\mathrm{b}$ (au) & $v_\mathrm{b}$ ($\kms$) & $dp$\\
\midrule
Edge & $103.4\pm 8.9$ & $2.45\pm 0.12$ & $0.37\pm 0.11$\\
Ridge & $123.0\pm 2.3$ & $1.86\pm 0.02$ & $0.67\pm 0.11$\\
\bottomrule
\end{tabular}
\tabnote{$v_\mathrm{b}$ is not corrected by the inclination factor of $\sin i$. \label{tab:best}}
\end{table}

\begin{table*}[t!]
\caption{Values calculated from the best-fit power-law function}
\centering
\begin{tabular}{ccccc}
\toprule
& $r$ range (au) & $v$ range ($\kms$) & $M_{\rm in}$ ($=M_b$; $\Msun$) & $M_{\rm out}$ ($\Msun$)\\
\midrule
Edge & 37.0 -- 185.5 & 1.48 -- 4.10 & 1.20$\pm$0.15 & 0.78$\pm$0.17\\
Ridge & 27.2 -- 157.5 & 1.39 -- 3.95 & 0.82$\pm$0.02 & 0.59$\pm$0.04\\
\bottomrule
\end{tabular}
\tabnote{$M_{\rm in}=M_b$ because $p_{\rm in}$ is fixed at 0.5 (Keplerian rotation). $M$ is inclination-corrected, while the $v$ range is not inclination-corrected. Note $M_{\rm out}$ no longer has a meaning of the central stellar mass because $dp > 0$. \label{tab:cal}}
\end{table*}

As another example, a case with \texttt{include\_pin=True} is performed, i.e., the inner power-law index is not fixed to the Keplerian law. The bast-fit parameters of the edge method are $(r_\mathrm{b}, v_\mathrm{b}, p_\mathrm{in}, dp)=(83.3\pm 9.8~\mathrm{au}, 2.82\pm 0.18~\kms, 0.41\pm 0.04, 0.37\pm 0.06)$. Those of the ridge method are $(123.9\pm 2.7~\mathrm{au}, 1.83\pm 0.03~\kms, 0.53\pm 0.01, 0.58\pm 0.12)$. The masses are calculated to be $(M_\mathrm{in}, M_\mathrm{b}, M_\mathrm{out})=(1.08\pm 0.24~\Msun, 1.27\pm 0.22~\Msun, 0.81\pm 0.21~\Msun)$ in the edge method, while they are $(0.87\pm 0.04~\Msun, 0.79\pm 0.03~\Msun, 0.59\pm 0.05~\Msun)$ in the ridge method.
When the error bars are included, the estimated $r_\mathrm{b}$, $v_\mathrm{b}$, and $dp$ are overall close to those estimated in the example with the fixed $p_\mathrm{in}$. The estimated inner index $p_\mathrm{in}$ is close to the Keplerian law, 0.5, in both methods, supporting $p_\mathrm{in}=0.5$ fixed in the previous example. 
In addition, the evidence is $\sim 0.8\times 10^{-15}$ for the edge points and $\sim 0.4\times 10^{-41}$ for the ridge points. These are $\sim 3$ times lower than those for the $p_\mathrm{in}$-fixed case, implying that the $p_\mathrm{in}$-fixed model could be better the $p_\mathrm{in}$-varied model.
When $p_\mathrm{in}$ is not exact 0.5, $M_\mathrm{in}$ and $M_\mathrm{b}$ are not exactly the same. In this case, a middle value between these masses will provide a reasonable estimate of the stellar mass ($M_*\sim 1.3$ and $0.8~\Msun$ in the edge and ridge methods, respectively), which is consistent with the example with the fixed $p_\mathrm{in}$. 

The fitting in the examples above used fixed values for the systemic velocity, the distance to the target, and the inclination angle. Among them, the systemic velocity can be incorporated in the fitting, if necessary, by setting {\tt include\_vsys=True}, while the distance and inclination angle need to be estimated outside our tool. When the uncertainties of the distance $d$ and the inclination angle $i$ (in the unit of radian) are $\Delta d$ and $\Delta i$, respectively, these cause additional uncertainties for estimated radius $r$, velocity $v$, mass $M$, and specific angular momentum $j$ as $\Delta r=r\Delta d/d$, $\Delta v=v\Delta i/\tan i$, $\Delta M=M\sqrt{4(\Delta i)^2/\tan ^2 i + (\Delta d)^2/d^2}$, and $\Delta j=j\sqrt{(\Delta i)^2/\tan ^2 i + (\Delta d)^2/d^2)}$. For example, when $\Delta d/d=0.05$, $\Delta i=3\arcdeg$, and $i=50\arcdeg$, $\Delta r/r=0.05$, $\Delta v/v=0.04$, $\Delta M/M=0.09$, and $\Delta j/j=0.06$.

\section{Discussion} \label{sec:discussion}
The example in the previous section demonstrates that our tool can provide a disk radius and a central stellar mass separately with the edge and ridge methods. The best-fit function including the outer power-law index also enables to calculate a specific angular momentum. The MCMC fitting provides a statistical error of each free parameter and the central stellar mass. In addition to the statistical error, the obtained disk radius and central stellar mass have uncertainties due to the difference between the edge and ridge methods. In the example presented in Section \ref{sec:act}, the difference between the estimated stellar masses by the two methods is larger than the statistical errors in the edge and ridge methods by a factor of $\sim20$ and $\sim3$, respectively. Thus, our tool adopts the edge and ridge values as upper and lower limits. Although previous works suggest how much the two methods over- or underestimates the central stellar mass \citep[e.g.,][]{mare20,aso15}, this is beyond the scope of our tool at this moment.

\subsection{Caveats} \label{subsec:caveats}
Lastly, we note caveats in the usage of our tool. First of all, how accurately the method in our tool estimates physical quantities ($M_*$, $R_{\rm disk}$, $p$, $j$, etc.) is still under heated debate, with other questions such as which of edge and ridge methods is better, whether these analytical methods are sufficient for a certain purpose or making a model with radiative transfer is necessary. The references in this report help to understand the current situation. In other words, it is currently difficult to evaluate any systematic error of the physical quantities. Our tool thus provides only a statistical error that originates in the given observational noise level. The statistical error is often smaller than differences due to choice of methods and parameters, such as edge versus ridge and $M_{\rm in}$ versus $M_b$. Hence, we emphasize that each free parameter could have uncertainties larger than the statistical error. The free parameters are selected on a case-by-case basis and could contribute to the uncertainty. For example, when $p_{\rm in}$ is not 0.5 but close to 0.5, the fitting provides different $M_{\rm in}$ and $M_b$. In comparison, the two masses are the same if $p_{\rm in}$ is fixed to be 0.5, which may be justified by theoretical prediction (an inner rotation must be in the Keplerian rotation). Hence, we recommend trying multiple parameter settings (single-power, double-power, fixing and varying $p_{\rm in}$ and $p_{\rm out}$, etc.) to check how robust a result is. For example, when a double-power fitting provides $r_\mathrm{b}$ close to the innermost or outermost radius, this result suggests that the single-power model is better than the double-power model in the given PV diagram. In such a double-power case, a power-law index tends to be strangely high or low simply because there are too few points on the inner ($r<r_\mathrm{b}$) or outer ($r>r_\mathrm{b}$) side. Similarly, when a double-power fitting provides $p_\mathrm{in}$ close to the Keplerian index, 0.5, such as the second example in Section \ref{sec:act}, it may be better to fix $p_\mathrm{in}=0.5$ because the Keplerian rotation is supported by physics, and fixing $p_\mathrm{in}$ may provide a better model in the sense of the evidence ratio, i.e., the Bayes factor.

The contribution of infall motion is also controversial. Although PV diagrams along the disk major axis are supposed to trace rotation in the system, the observed velocity is the line-of-sight component of a combination of rotation and infall motions. The outer power-law index is introduced based on a picture where gas is infalling and thus following a lower power-law index in an outer part than in an inner part. Because the infall velocity is expected to be proportional to $r^{-0.5}$, as is the case with free fall, and is more dominant than rotation in outer radii, the infall motion could cause $p_{\rm out}$ to be closer to 0.5. Whether an infall motion is dominant can be verified by using a 2D distribution of the mean velocity (moment 1) and checking the quadrants in a PV diagram where emission is not expected in the case only with rotation. When the observed system includes infalling gas, the PV diagram along the disk major axis has emission in all the four quadrants, which could shift the edge and ridge points from the radius/velocity in the case only with rotation. In such a case, it is worthwhile to try limiting the velocity range to high velocities where emission is seen only in two quadrants to focus on the rotation-dominant part.

\section{Summary}

We have developed the \texttt{pvanalysis} tool, which is implemented in a python library \texttt{SLAM}. The main objective of this tool is to kinematically identify disks around YSOs in the protostellar phase, which are embedded in infalling envelopes, by analyzing PV diagrams of emission line data. As presented in Section \ref{sec:methodologies} and \ref{sec:usage}, the straightforward concept and usage of this tool make it highly accessible to examine Keplerian rotation of the disks. In Section \ref{sec:act}, we demonstrated the application of this tool to real observational data, which highlights its ability to extract key observational features and to distinguish between the rotational motions of a disk and an infalling envelope. It is important to acknowledge that certain uncertainties still exist, which are currently not taken into account in the tool. These include systemic errors inherent in the method itself and potential contamination from infalling velocities, as discussed in Section \ref{sec:discussion}. Nevertheless, the analytic approach of the \texttt{pvanalysis} tool with few assumptions and its simple usage would be a great asset for identifying disks in the protostellar phase and measuring the dynamical mass of central objects.


\acknowledgments

This paper makes use of the following ALMA data: ADS/JAO.ALMA\#2013.1.01086.S and ADS/JAO.ALMA\#2015.1.01415.S. ALMA is a partnership of ESO (representing its member states), NSF (USA) and NINS (Japan), together with NRC (Canada), MOST and ASIAA (Taiwan), and KASI (Republic of Korea), in cooperation with the Republic of Chile. The Joint ALMA Observatory is operated by ESO, AUI/NRAO and NAOJ.

\bibliographystyle{custom}
\bibliography{sample,ref_sai}

\end{document}